\documentclass[runningheads]{llncs}
\usepackage{graphicx}
\usepackage{color}
\usepackage{multirow}
\usepackage{tabularx}
\usepackage{url}
\PassOptionsToPackage{hyphens}{url}\usepackage{hyperref}

\begin{document}
\title{A Comparison of Question Rewriting Methods for Conversational Passage Retrieval}
\titlerunning{A Comparison of Question Rewriting Methods}
%
\author{Svitlana Vakulenko\inst{1} \and
Nikos Voskarides\inst{1} \and
Zhucheng Tu\inst{2} \and
Shayne Longpre\inst{2}}

%
\institute{University of Amsterdam \and Apple Inc.}

\authorrunning{Vakulenko et al.}
\maketitle              
\begin{abstract}
Conversational passage retrieval relies on question rewriting to modify the original question so that it no longer depends on the conversation history.
Several methods for question rewriting have recently been  proposed, but they were compared under different retrieval pipelines.
We bridge this gap by thoroughly evaluating those  question rewriting methods on the TREC CAsT 2019 and 2020 datasets under the same retrieval pipeline.
We analyze the effect of different types of question rewriting methods on retrieval performance and show that by combining question rewriting methods of different types we can achieve state-of-the-art performance on both datasets.\footnote{Resources can be found at \url{https://github.com/svakulenk0/cast_evaluation}.}

\end{abstract}
\section{Introduction}

Conversational search aims to provide automated support for natural and effective human–information interaction~\cite{DBLP:journals/dagstuhl-reports/AnandCJSS19}.
The TREC Conversational Assistance Track (CAsT) introduced the task of conversational (multi-turn) passage retrieval (PR)~\cite{dalton2019trec}, where the goal is to retrieve  short passages of text from a large passage collection that answer the information need at the current turn. 

One prominent challenge in conversational PR is that the question at the current turn often requires information from the conversation history (questions and passages retrieved in previous turns) to be interpreted correctly.
A proposed solution to this challenge is question rewriting (or resolution, QR), i.e., modifying the question such that it no longer depends on the conversation history.
For instance, the question ``What did he work on?'' can be rewritten into ``What did Bruce Croft work on?'' based on the conversation history (see Table~\ref{samples} for the complete example).

Recently proposed methods for QR in conversational PR can be categorized into two types, namely sequence generation and term classification.
Sequence generation QR methods generate natural language sequences using the conversation history~\cite{DBLP:journals/corr/abs-2004-14652,DBLP:conf/sigir/YuLYXBG020}, while term classification QR methods add terms from the conversation history to the current turn question~\cite{DBLP:conf/sigir/MeleMN0TF20,DBLP:conf/sigir/VoskaridesLRKR20}.
The former can be trained using human generated rewrites or data obtained from search sessions and heuristics~\cite{DBLP:journals/corr/abs-2004-14652,DBLP:conf/sigir/YuLYXBG020}, while the latter are either heuristic-based~\cite{DBLP:conf/sigir/MeleMN0TF20}, or trained using human generated rewrites or distant supervision~\cite{DBLP:conf/sigir/VoskaridesLRKR20}.

In this paper, we conduct a systematic evaluation of the state-of-the-art QR methods under the same retrieval pipeline on the CAsT 2019 and 2020 datasets. 
While CAsT 2019 only depends on the previous questions in the conversation, CAsT 2020 also includes questions that depend on the previously retrieved passages.
Our results provide insights on the ability of the QR methods to account for the conversation history, as well as on the potential of combining QR methods of different types for improving retrieval effectiveness.
\section{Task Definition}
We model the conversational PR task as a sequence of two subtasks: (1) question rewriting (QR) and (2) passage retrieval (PR)~\cite{DBLP:journals/corr/abs-2004-14652,DBLP:conf/sigir/VoskaridesLRKR20,DBLP:conf/sigir/YuLYXBG020}.
In this paper, we focus on the QR subtask and investigate the impact of QR on PR performance.

In the QR subtask, we are given the current turn question $Q_i$ and a sequence of question-answer pairs $H := Q_1, A_1, \ldots, Q_{i-1}, A_{i-1}$  (the conversation history).
The current turn question $Q_i$ may depend on the conversation history $H$ and thus some information in $H$ is required to correctly interpret $Q_i$. 
The goal of QR is to generate a question rewrite $Q'_i$ that no longer depends on $H$.

In the PR subtask, we are given the question rewrite $Q'_i$ and a passage collection $C$, and the goal is to retrieve a list of passages $R$ sorted by their relevance to $Q'_i$ from $C$.
If $Q'_i$ is semantically equivalent to $\langle Q_i, H \rangle$, we expect $R$ to constitute relevant passages for $\langle Q_i, H \rangle $.

\section{Experimental Setup}
We aim to answer the following research questions:

\noindent\textbf{RQ1} How do different QR methods perform on the two datasets we consider (CAsT 2019 and CAsT 2020)?

\noindent\textbf{RQ2} Can we combine different QR models to improve retrieval performance?

Following previous work, we perform both intrinsic and extrinsic evaluation~\cite{DBLP:journals/corr/abs-2010-04898,DBLP:conf/sigir/VoskaridesLRKR20}. 
In intrinsic evaluation, we compare  rewrites produced by QR methods with manual rewrites produced by human annotators using ROUGE-1 Precision (P), Recall (R) and F-measure (F)~\cite{DBLP:journals/corr/abs-2010-04898}.
\footnote{We use ROUGE-1 to measure unigram overlap after punctuation removal, lower casing and Porter stemming. We use the following ROUGE implementation: \url{https://github.com/google-research/google-research/tree/master/rouge}}
In extrinsic evaluation, we measure PR performance when using different QR methods 
using standard ranking metrics: NDCG@3, MRR and Recall@1000.

\subsection{Question rewriting methods}
\label{qr_models}

We compare the following question rewriting methods:
\begin{itemize}
    \item \textbf{Original} The original current turn question without any modification.

    \item \textbf{Human} The gold standard rewrite of the current turn question produced by a human annotator.

    \item \textbf{Rule-Based} and \textbf{Self-Learn} model question rewriting as a sequence generation task and use GPT-2 to perform generation~\cite{DBLP:conf/sigir/YuLYXBG020}.
    In order to gather training data, these methods  convert ad-hoc search sessions to conversational search sessions either by using heuristic rules (\textbf{Rule-Based}) or by using self-supervised learning (\textbf{Self-Learn}).
    
    \item \textbf{Transformer++}~\cite{DBLP:journals/corr/abs-2004-14652} is a GPT-2 sequence generation  model. It was trained on CANARD, a conversational question rewriting dataset~\cite{elgohary2019can}.

    \item \textbf{QuReTeC}~\cite{DBLP:conf/sigir/VoskaridesLRKR20} models question rewriting as term classification, i.e., predicting which terms from the conversation history to add to the current turn question.
    It uses BERT to perform term classification and can be trained using human rewrites or distant supervision obtained from query-passage relevance labels.
    In this paper, we use the model trained on CANARD~\cite{elgohary2019can} to be comparable with \textbf{Transformer++}. Since \textbf{QuReTeC} does not generate natural language text but rather appends a bag-of-words (BoW) to the original question, we also introduce an oracle \textbf{Human-BoW} as an upper-bound for \textbf{QuReTeC} performance.

\end{itemize}

\subsection{Datasets}
\begin{table}[t]
\centering
\caption{Datasets statistics.}\label{datasets}
\begin{tabularx}{0.63\textwidth}{l@{\hskip .15in}c@{\hskip .15in}c@{\hskip .15in}cr}
\hline
\multicolumn{1}{c}{Dataset} & \#Topics & \#Questions & \#Copy & (\%)  \\
\hline
CAsT 2019 & 50 & 479 & 88 & (21) \\
CAsT 2020 & 25 & 216 & 5 & (3) \\
\hline
\end{tabularx}
\end{table}
We use the recently constructed TREC CAsT 2019 and CAsT 2020 datasets~\cite{dalton2019trec}.
Table~\ref{datasets} shows basic statistics of the datasets.
\textbf{Copy} indicates the number of questions for which the human rewrite is exactly the same as their corresponding original question.
This statistic shows that in contrast to CAsT 2019, in CAsT 2020, only a very few questions can be copied verbatim and the majority of questions require extra terms.

Another major difference between the two datasets is that the current turn question in CAsT 2020 may also depend on the answer passage to  the previous turn question ($A_{i-1}$), while in CAsT 2019 the current turn question depends only on the questions of the previous turns in the conversation history ($Q_1, Q_2, \ldots, Q_{i-1}$).
Therefore, we experiment with two variations of input to the QR models: (1) all previous questions (indicated as \textbf{Q}) and (2) all previous questions and the answer passage to the previous turn question (indicated as \textbf{Q\&A}).\footnote{We use the answer passage to the previous turn question retrieved by the \emph{automatic} rewriting system provided by the TREC CAsT 2020 organizers.}

\subsection{Passage retrieval pipeline}
All QR methods described in Section \ref{qr_models} were previously evaluated on CAsT 2019 using different retrieval pipelines.
For a fair comparison, we evaluate the QR methods on both CAsT 2019 and CAsT 2020 using the same passage retrieval pipeline.

We use a standard two-stage pipeline for passage retrieval, consisting of an unsupervised ranker for initial retrieval performing efficient lexical match (BM25) and a supervised reranker (BERT) over the top-1000 passages returned by initial retrieval~\cite{nogueira2019passage}.\footnote{Note that our pipeline  outperforms the official baseline provided by the TREC CAsT organizers for both 2019 and 2020 datasets for all query rewriting methods they considered. Since our focus is on comparing different query rewriting methods, we do not report those results for brevity.}
Both components were fine-tuned on a subset of the MS MARCO dataset ($k_1=0.82, b=0.68$).\footnote{\url{https://github.com/nyu-dl/dl4marco-bert}}

\section{Results}

\subsection{QR methods comparison}
Here we answer \textbf{RQ1}: How do different QR methods perform on the two datasets we consider?

\begin{table}[t]
\centering
\caption{Evaluation of question rewriting methods on  CAsT 2019.}\label{2019}
\begin{tabular}{lcccccc}
\hline
\multicolumn{1}{c}{\multirow{2}{*}{QR Method}} &
\multicolumn{1}{c}{Recall@1000} & \multicolumn{2}{c}{NDCG@3} & \multicolumn{3}{c}{ROUGE-1} \\
\multicolumn{1}{c}{} & Initial & Initial & Reranked & \multicolumn{1}{c}{P} & \multicolumn{1}{c}{R} & \multicolumn{1}{c}{F} \\
\hline
Original & 0.417 & 0.131 & 0.266 & 0.92 & 0.76 & 0.82 \\
Transformer++ Q & 0.743 & 0.265 & \bf 0.525 & \bf 0.96 & 0.88 & \bf 0.91 \\
Self-Learn Q & 0.725 & 0.261 & 0.513 & 0.93 & 0.89 & 0.90 \\
Rule-Based Q & 0.717 & 0.248 & 0.487 & 0.94 & 0.89 & 0.91 \\
QuReTeC Q & \bf 0.768 & \bf 0.296 &  0.500 & 0.89 & \bf 0.90 & 0.89 \\
\hline
Transformer++ Q + QuReTeC Q & \bf 0.791 & 0.300 & \bf 0.546 & \bf 0.93 & 0.91 & \bf 0.91 \\
Self-Learn Q + QuReTeC Q & 0.785 & 0.293 & 0.519 & 0.90 & \bf 0.93 & \bf 0.91 \\
Rule-Based Q + QuReTeC Q & 0.783 & \bf 0.301 & 0.534 & 0.91 & \bf 0.93 & \bf 0.91 \\
\hline
Human-BoW Q & 0.769 & 0.297 & 0.524 & 0.91 & 0.90 & 0.90 \\
Human & 0.803 & 0.309 & 0.577 & 1.00 & 1.00 & 1.00 \\
\hline
\end{tabular}
\end{table}

\textbf{CAsT 2019.} 
In Table~\ref{2019}, we observe that QuReTeC outperforms all other methods in initial retrieval (Recall@1000 and NDCG@3).
However, we see that Transformer++ Q outperforms QuReTeC in reranking (NDCG@3).
This may indicate that the reranking component (BERT) is more sensitive to rewritten questions that do not resemble natural language text (produced by QuReTeC) than the initial retrieval component (BM25).
This is also reflected in the ROUGE-1 metric variations:
ROUGE-1 R is generally in agreement with initial retrieval performance.
This is expected since our initial retrieval component is BoW and does not get substantially affected by missing or incorrect terms such as pronouns and stopwords, which are usually insignificant for lexical matching (see Human-BoW in Table~\ref{2019}). 
ROUGE-1 P, however, favours the sequence generation methods, and penalizes QuReTeC, since QuReTeC does not have a mechanism to delete or replace such terms from the original question.

\begin{table}[t]
\centering
\caption{Evaluation of question rewriting methods on  CAsT 2020.}\label{2020}
\begin{tabular}{lcccccc}
\hline
\multicolumn{1}{c}{\multirow{2}{*}{QR Method}} & \multicolumn{1}{c}{Recall@1000} & \multicolumn{2}{c}{NDCG@3} & \multicolumn{3}{c}{ROUGE-1} \\
\multicolumn{1}{c}{} & Initial & Initial & \multicolumn{1}{c}{Reranked} & \multicolumn{1}{c}{P} & \multicolumn{1}{c}{R} & \multicolumn{1}{c}{F} \\
\hline
Original & 0.251 & 0.068 & 0.193 & \bf 0.87 & 0.66 & 0.74 \\
Transformer++ Q\&A & 0.351 & 0.098 & 0.252 & 0.75 & 0.69 & 0.70 \\
Self-Learn Q\&A & 0.462 & 0.156 & 0.342 & 0.84 & 0.73 & 0.76 \\
Rule-Based Q\&A & 0.455 & 0.137 & 0.339 & 0.84 & 0.75 & \bf 0.78 \\
QuReTeC Q\&A & \bf 0.531 & \bf 0.171 & \bf 0.370 & 0.82 & \bf 0.77 & \bf 0.78 \\
\hline
Transformer++ Q + QuReTeC Q\&A & 0.525 & 0.160 & 0.351 & \bf 0.83 & 0.77 & 0.78 \\
Self-Learn Q + QuReTeC Q\&A & \bf 0.567 & 0.168 & \bf 0.375 & 0.82 & \bf 0.79 & \bf 0.79 \\
Rule-Based Q\&A + QuReTeC Q\&A & 0.519 & \bf 0.173 & 0.362 & 0.80 & \bf 0.79 & 0.78 \\
\hline
Human-BoW Q & 0.579 & 0.189 & 0.465 & 0.89 & 0.81 & 0.84 \\
Human-BoW Q\&A & 0.649 & 0.226 & 0.465 & 0.88 & 0.85 & 0.86 \\
Human & 0.707 & 0.240 & 0.531 & 1.00 & 1.00 & 1.00 \\
\hline
\end{tabular}
\end{table}

\medskip

\textbf{CAsT 2020.} In Table~\ref{2020}, we observe that the retrieval performance of Original and Human is much lower than in Table~\ref{2019}, which indicates that CAsT 2020 is more challenging than CAsT 2019.\footnote{Recall that questions in CAsT 2020 may depend on the answer of the previous turn question, but this is not the case in CAsT 2019.}
We observe that QuReTeC outperforms all other methods in all ranking metrics.
This indicates that QuReTeC better captures relevant terms both from the previous turn questions and the answer passage to the previous turn question than the other QR methods.
Similarly to Table~\ref{2019}, ROUGE-1 R is in agreement with initial retrieval performance.
As for ROUGE-1 P, we observe that it is not as important for retrieval as in Table~\ref{2019}.
\begin{figure}[t]
\centering
\includegraphics[scale=0.35]{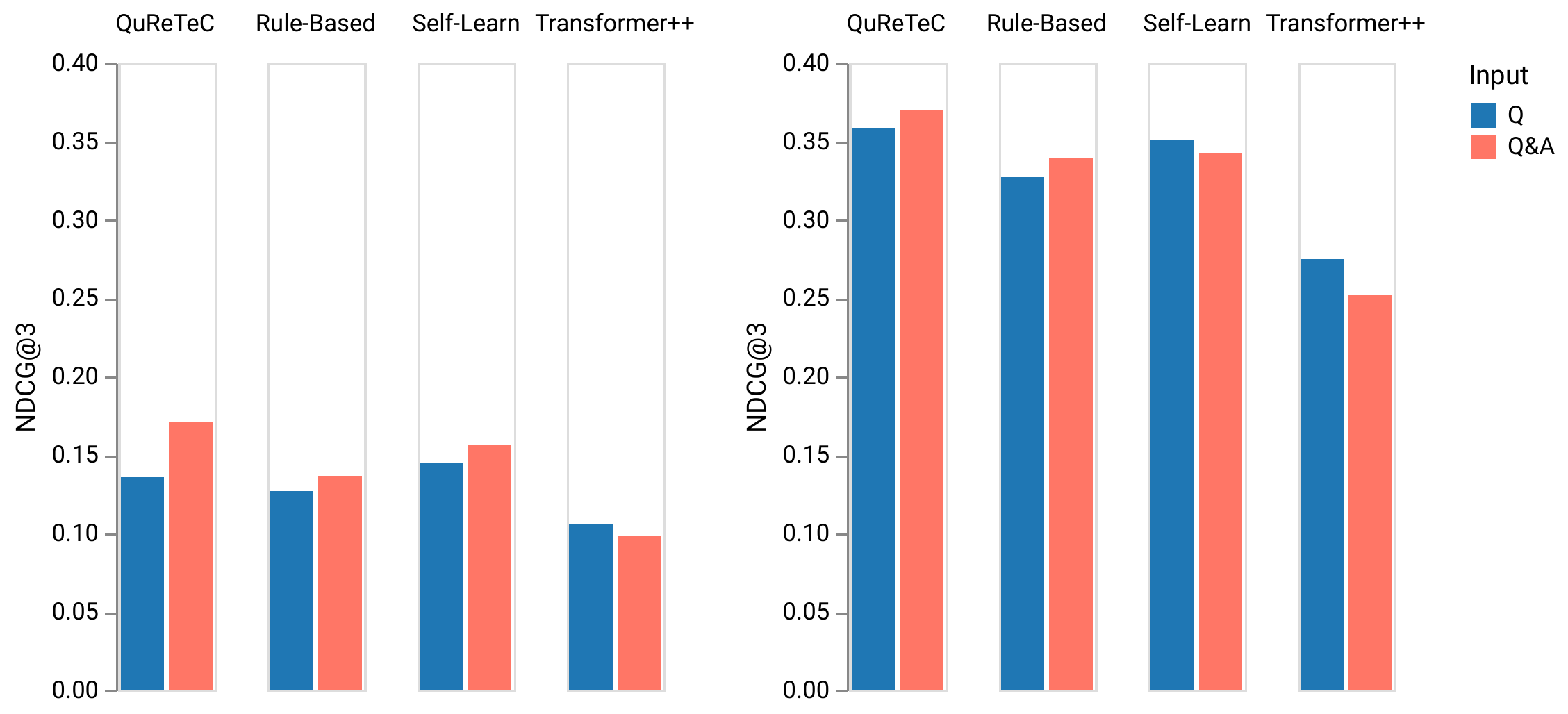}

\caption{Initial retrieval (left) and reranking (right) performance on CAsT 2020 when the answer passage to the previous turn question is used (Q\&A) or not used (Q) as input to the QR methods.
}
\label{qna}
\end{figure}
Next, we assess the contribution of the answer passage to the previous turn question on QR performance.
In Figure~\ref{qna}, we observe that most QR methods (except Transformer++) do benefit from using the answer passage, with QuReTeC having the biggest gain in initial retrieval.
\begin{table}[t]
\caption{Example question rewrites for the topic in CAsT 2020 starting with ``Who are some of the well-known Information Retrieval researchers?''.
}\label{samples}
\begin{tabularx}{\textwidth}
{p{0.23\textwidth}@{\hskip .15in} p{0.22\textwidth}@{\hskip .15in} p{0.22\textwidth}@{\hskip .15in} p{0.22\textwidth}}
\hline
\multicolumn{1}{c}{Answer Passage} & \multicolumn{1}{c}{Original} & \multicolumn{1}{c}{Rule-Based Q\&A} & \multicolumn{1}{c}{QuReTeC Q\&A} \\
\hline
Bruce Croft formed the Center ... & What did he work on? & What did Bruce Croft work on? & What did he work on? croft bruce \\ 
\hline
Karpicke and Janell R. Blunt (2011) followed up ... & Who are some important British ones? & Who are some important British ones? & Who are some important British ones? information retrieval \\
\hline
\end{tabularx}

\end{table}
Table~\ref{samples} shows examples of question rewrites produced by Rule-Based and QuReTeC.

\subsection{Combining QR methods}

Next we answer \textbf{RQ2}: Can we combine different QR models to improve performance?
In order to explore whether combining QR methods of different types (sequence generation or term classification) can be beneficial, we simply append terms from the conversation history predicted as relevant by QuReTeC to the rewrite produced by one of the sequence generation methods.
We found that by doing this we can improve upon individual QR methods and achieve state-of-the-art retrieval performance on CAsT 2019 by combining Transformer++ Q with QuReTeC Q (see Table~\ref{2019}), and on CAsT 2020 by combining Self-Learn Q and QuReTeC Q\&A (see Table~\ref{2020}); however the gains on CAsT 2020 are smaller.

\section{Conclusion}
We evaluated alternative question rewriting methods for conversational passage retrieval on the CAsT 2019 and CAsT 2020 datasets.
On CAsT 2019, we found that QuReTeC performs best in terms of initial retrieval, while Transformer++ performs best in terms of reranking.
On CAsT 2020, we found that QuReTeC performs best both in terms of initial retrieval and reranking.
Moreover, we achieved state-of-the-art ranking performance on both datasets using a simple method that combines the output of QuReTeC (a term classification method) with the output of a sequence generation method.
Future work should focus on developing more advanced methods for combining term classification and sequence generation question rewriting methods.

\noindent\textbf{Acknowledgements} We thank  Raviteja Anantha for providing the rewrites of the Transformer++ model.

\clearpage

\bibliographystyle{splncs04}
\bibliography{refs}
    
\end{document}